\documentclass[useAMS,usenatbib]{mn2e}

\usepackage[dvips]{graphicx}
\usepackage[]{txfonts}
\def\simless{\mathbin{\lower 3pt\hbox
{$\rlap{\raise 5pt\hbox{$\char'074$}}\mathchar"7218$}}}   %< or of order
\def\simmore{\mathbin{\lower 3pt\hbox
{$\rlap{\raise 5pt\hbox{$\char'076$}}\mathchar"7218$}}}   %> or of order
\def\Msun{{\rm M}_\odot}                                       % solar masses 
\newcommand{\be}{\begin{equation}}
\newcommand{\ee}{\end{equation}}
\title{Fast TeV variability in blazars: jets in a jet}

\author[Dimitrios Giannios, Dmitri A. Uzdensky and Mitchell C. Begelman]
{Dimitrios Giannios$^{1,2}$\thanks{E-mail: giannios@astro.princeton.edu
 (DG)}, Dmitri A. Uzdensky$^{1}$ and  Mitchell C. Begelman$^{3,4}$\\
$^{1}$Department of Astrophysical Sciences, Peyton Hall, Princeton
  University, Princeton, NJ 08544, USA\\
$^{2}$Max Planck Institute for Astrophysics, Box 1317, D-85741 Garching,
  Germany\\
$^{3}$Joint Institute for Laboratory Astrophysics, University of Colorado,
  Boulder, CO 80309, USA\\
$^{4}$Department of Astrophysical and Planetary Sciences, University of
  Colorado, Boulder}

\begin{document}
\date{Received / Accepted}
\pagerange{\pageref{firstpage}--\pageref{lastpage}} \pubyear{2009}

\maketitle

\label{firstpage}

\begin{abstract}
The fast TeV variability of the blazars Mrk 501 and PKS 2155--304 implies a
compact emitting region that moves with a bulk Lorentz factor of
$\Gamma_{\rm em}\sim 100$ toward the observer. The Lorentz
factor is clearly in excess of the jet Lorentz factors $\Gamma_{\rm j}\simless 10$
measured on sub-pc scales in these sources. We propose that the TeV
emission originates from compact emitting 
regions that move relativistically {\it within} a jet of bulk $\Gamma_{\rm j}\sim$
10. This can be physically realized in 
a Poynting flux-dominated jet. We show that if a large fraction of the luminosity 
of the jet is prone to magnetic dissipation through reconnection, then material 
outflowing from the reconnection regions can efficiently power the observed TeV
flares through synchrotron-self-Compton emission. The model predicts
simultaneous far UV/soft X-ray flares.
\end{abstract} 
  

\section{Introduction} 
\label{intro}

There are two cases of blazars (Mrk 501 and PKS 2155--304) with flaring TeV emission that varies
on timescales of 3-5 minutes (Albert et al. 2007; Aharonian et al. 2007). 
This variability timescale is much shorter than the
light crossing time of the gravitational radius $t_{lc}\sim$hours of the supermassive black
holes of these blazars (for inferred masses $M_{\rm BH}\sim 10^9\Msun$). This implies 
a very compact $\gamma$-ray emitting region. Furthermore, the fact that the TeV photons
escape the production region implies that the emitting plasma 
moves with bulk $\Gamma_{\rm em}\simmore 50$ so as to avoid pair creation
through interaction with soft radiation fields\footnote{It is possible to
relax the transparency constraint on the bulk Lorentz factor of the emitting
region by supposing an extremely sharp lower
cutoff in the electron distribution (Boutelier, Henri \& Petrucci 2008).} 
(Begelman, Fabian \& Rees 2008; Mastichiadis \& Moraitis 2008). It therefore appears
natural that a compact region within the blazar jet is the source of the TeV flares.

On the other hand, the jets of PKS 2155--304 and Mrk 501 have been resolved 
on sub-pc scales and show patterns that move with moderate Lorentz factor 
$\Gamma_{\rm  j}\simless 10$ (Piner \& Edwards 2004; Giroletti et al.
2004), much less than the one needed for TeV $\gamma$-rays
to escape. This apparent contradiction can be avoided if, for some reason, the
jet is efficiently decelerated on sub-pc scales (Georganopoulos \&
Kazanas 2003;  Levinson 2007) after the TeV emission has taken place.
The deceleration may be the result of radiative feedback in a spine/layer
configuration (Ghisellini, Tavecchio \& Chiaberge 2005; Tavecchio \&
Ghisellini 2008). Alternatively, finite opening angle effects in a
$\Gamma_{\rm j}\sim 50$ jet can result in slow radio knot motions in 
blazars (Gopal-Krishna, Dhurde \& Wiita 2004; Gopal-Krishna et al. 2007).

Here, we propose an alternative explanation for the origin of the TeV
emission as a ``jets-in-a-jet'' model. We argue that compact emitting 
regions that move relativistically {\it within} a jet of bulk $\Gamma_{\rm
 j}\sim 10$ can power the TeV flares. Ghisellini et al. (2009) proposed a
similar basic idea of internal jet motions.  We provide a different 
context/cause for these motions, arguing that they can be physically realized in 
a Poynting flux-dominated flow (PDF). We show that, if a large fraction of the luminosity 
of a PDF is occasionally prone to magnetic dissipation through reconnection,
then material outflowing at relativistic speed from the reconnection regions 
can efficiently power the observed TeV flares.

\section{Jets-in-a-jet model}

Consider a jet that moves radially with bulk $\Gamma_{\rm j}$, containing a 
blob of plasma with a characteristic Lorentz factor
$\Gamma_{\rm co}$ at angle $\theta'$ with respect to the 
radial direction (measured in the jet rest frame). 
All primed/tilded quantities are measured in the rest frame of the jet/blob respectively. 
In the lab frame the blob moves with
\be
\Gamma_{\rm em}=\Gamma_{\rm j} \Gamma_{\rm co}(1+v_jv_{\rm co}\cos \theta')
\ee 
and at angle 
\be
\tan\theta=\frac{v_{\rm co}\sin\theta'}{\Gamma_{\rm j}(v_{\rm
    co}\cos\theta'+v_{\rm j})}
\label{angle}
\ee
with respect to the radial direction. 
For a large range of angles of order $\theta'\sim \pi/2$, expressions
(1) and (2) give $\Gamma_{\rm em}\sim \Gamma_{\rm j} \Gamma_{\rm co}$ and 
$\theta\sim 1/\Gamma_{\rm j}$. The blob moves with 
$\Gamma_{\rm em}\gg \Gamma_{\rm j}$ provided 
that the motions within the jet are relativistic. 
Such fast internal motions are possible in a Poynting flux-dominated flow where 
MHD waves approach the speed of light.

\subsection{The jet}
\label{}

For more quantitative estimates we consider a jet with (isotropic) luminosity $L_{\rm j}$  
that moves with the  bulk $\Gamma_{\rm j}$. The jet is assumed to be strongly
magnetized with Poynting-to-kinetic flux ratio (magnetization) $\sigma\gg 1$. 
As reference values, we use $\Gamma_{\rm j}=10$
and $\sigma=100$. The Poynting luminosity of the jet may be  
inferred from the flaring isotropic luminosity of PKS
2155--304 and is set to $L_{\rm j}= 10^{47}$ erg/s. 

The energy density in the jet is (as measured in a frame comoving with the jet)
\be
e'_{\rm j}=L_{\rm j}/4\pi r^2c\Gamma_{\rm j}^2=12 L_{\rm
  j,47}r_2^{-2}\Gamma_{\rm j,1}^{-2}\quad \rm{erg/cm^3},
\ee
where  $A=10^xA_x$ and the spherical radius is $R=r R_{\rm g}$ with 
$R_{\rm g}=1.5\times 10^{14}$ cm, corresponding to the gravitational radius of a
black hole of $10^9$ solar masses.  
The magnetic field strength in the jet is
\be
B'_{\rm j}=\sqrt{4\pi e'_{\rm j}}=12  L_{\rm j,47}^{1/2}r_2^{-1}
\Gamma_{\rm j,1}^{-1}\quad {\rm Gauss}.\ee
For a proton-electron jet, the particle number density in the jet is
\be
n'_{\rm j}=B_{\rm j}^2/4\pi c^2\sigma m_{\rm p}=80 L_{\rm
  j,47}r_2^{-2}\Gamma_{\rm j,1}^{-2}\sigma_2^{-1}\quad
{\rm cm^{-3}}.
\ee

\subsection{The emitting blob}

We assume that a fraction of the magnetic energy of the jet 
is occasionally dissipated through reconnection. In the PDF considered here,
current-driven instabilities are the most relevant ones in triggering the
dissipation (e.g. Eichler 1993; Begelman 1998; Giannios \& Spruit 2007; see,
however, McKinney \& Blandford 2009). Alternatively,
reversals in polarity of the magnetic field that threads the black hole can 
lead to magnetic reconnection in the jet (see also Sect.~\ref{sec-conclusions}).

Our picture for relativistic reconnection is the following (Lyubarsky 2005).
High-$\sigma$ material is advected into the reconnection region where the release
of magnetic energy takes place.  Part of the dissipated magnetic energy 
serves to give bulk acceleration of the ``blob'' (in the rest frame of the jet) and the
rest to heat the outflowing material to relativistic temperature.
We explore the possibility that emission from the outflowing material produces the 
TeV flares and we refer to it as the ``emitting blob'' or simply ``blob'' (see
Fig.~1).

For our quantitative estimates that follow, we adopt the relativistic generalization of 
Petschek-type reconnection worked out by Lyubarsky (2005; see also Watanabe \&
Yokoyama 2006 for relativistic MHD simulations that support this picture).
In this model, the material leaves the reconnection region with bulk
$\Gamma_{\rm co}$ close to the Alfv\'en speed of the upstream plasma
$\Gamma_{\rm co}\sim \sqrt {\sigma}\simeq 10\sigma_2^{1/2}$ in the 
rest frame of the jet (Petschek 1964; Lyutikov \& Uzdensky 2003; Lyubarsky
2005). For the last expression to be valid,
we assume that the guide field (i.e. non-reversing field component) is not strong
enough to affect the reconnection dynamics (i.e. $B'_{\rm guide}\simless
B'_{\rm j}/\sqrt{\sigma}$; see also Sec.~5 for when this condition may be satisfied). 
As seen in the lab frame, plasma is ejected from the reconnection region with
$\Gamma_{\rm em}\sim\Gamma_{\rm j}\Gamma_{\rm co}=100\Gamma_{\rm
  j,1}\sigma_2^{1/2}$.  The ratio of the thermal energy to rest mass in the 
blob frame is $\tilde{e}_{\rm em}/ \tilde{\rho}_{\rm em} c^2 \sim \sqrt {\sigma}$ and
reconnection leads to compression of the outflowing material $\tilde{\rho}_{\rm
  em}\sim \sqrt{\sigma}\rho'_{\rm j}$. 
The energy density in the blob is  (Lyubarsky 2005)
\be
\tilde{e}_{\rm em}\sim \sqrt{\sigma}\tilde{\rho}_{\rm em}c^2\sim \sigma
  \rho'_{\rm j} c^2=12 L_{\rm j,47}r_2^{-2}\Gamma_{j,1}^{-2}\quad
\rm{erg/cm^3}.
\label{eblob}
\ee
The fact that this is similar to eq.~(3) is just a consequence
of the pressure balance across the reconnection region.

Even though we consider a PDF jet, the emitting (downstream) region is not
necessarily magnetically dominated since a large part of the magnetic energy
dissipates in the reconnection region. This has important implications for the 
radiative processes discussed below.    
On the other hand, the blob material may remain strongly magnetized. 
Any guide field in the reconnection region will be amplified by compression
and will not dissipate. Lyubarsky (2005) shows that for a guide field  
$B'_{\rm guide}\simless B'_{\rm j}/\sqrt{\sigma}$, the magnetization of the 
blob (downstream plasma) is $\sigma_{\rm em }\simless 1$. The magnetic field in the
blob rest frame is roughly estimated to be
\be
\tilde{B}_{\rm em}\simless \sqrt{4\pi \tilde{e}_{\rm em}}=12  
L_{\rm j,47}^{1/2}r_2^{-1}\Gamma_{j,1}^{-1}\quad
{\rm Gauss}.
\ee

If electrons receive an appreciable fraction of the released energy $f\sim
0.5$, they are heated to characteristic \be \gamma_e\sim f
\sqrt{\sigma}m_p/m_e\sim 10^4f_{1/2}\sigma_2^{1/2}, \ee assumed to be
isotropic in the blob rest frame.

%-----------------------------------------------------------------  
\begin{figure}
\resizebox{\hsize}{!}{\includegraphics[]{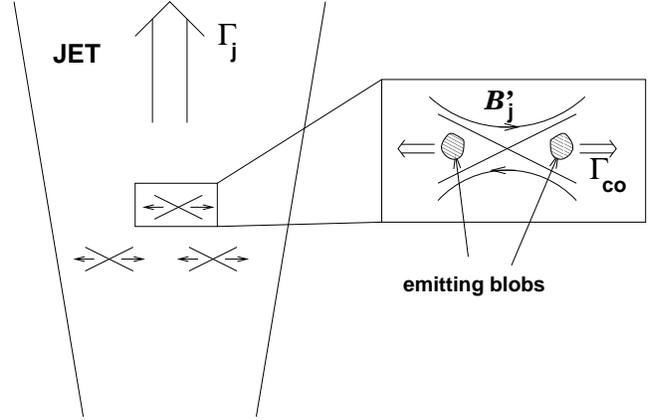}}
\caption[] {Schematic representation of the geometry of the 
``jets in a jet'' {\it shown in a frame comoving with the jet}. 
Right: the reconnection region enlarged. Plasma heated and 
compressed by  magnetic reconnection leaves the 
reconnection region at relativistic speed $\Gamma_{\rm co}\gg 1$ 
within the jet in the form of blobs. Each blob emits efficiently
through synchrotron-self-Compton in a narrow beam within the 
jet emission cone, powering a fast evolving soft X-ray and TeV flare. 
The sequence of flares seen in PKS 2155--304 may be the result of 
multiple reconnection regions or 
intrinsic instabilities (e.g., tearing) of one large reconnection region.   
\label{fig1}}
\end{figure}
%----------------------------------------------------------------- 

\subsubsection{The blob size}

From the observed energy of the TeV flares, we can estimate the 
energy contained in each blob. Combined with the energy density (\ref{eblob}), 
we derive an estimate of the size of the blob.  

The TeV flares have observed (isotropic equivalent) luminosity 
$L_{\rm f}\sim 10^{47}$ erg/s (allowing for a few  times the observed 
energy to be emitted below $\sim 200$ GeV, the low energy
threshold of the observations) and duration of $t_{\rm f}\sim 300$ s.
The associated energy is then $E_{\rm f}=L_{\rm f}\times t_{\rm f}\simeq 3\times
10^{49}L_{\rm f,47} t_{\rm f,300}$ erg. 

In the model discussed here, the source of the flare moves with a bulk
$\Gamma_{\rm em}\gg 1$. Its emission is concentrated in a cone that corresponds
to a fraction $\sim 1/4\Gamma_{\rm em}^2=2.5\times
10^{-5}\Gamma_{\rm j,1}^{-2}\sigma_2^{-1}$ of the sky. The lab-frame energy 
(corrected for collimation) radiated from the blob is, thus, 
$E_{\rm rad}=E_{\rm f}/4\Gamma_{\rm em}^2\simeq 7.5\times 10^{44}
L_{\rm f,47} t_{\rm f,300}\Gamma_{\rm j,1}^{-2}\sigma_2^{-1}$ erg. The
 energy contained in the blob is
$E_{\rm em}=E_{\rm rad}/f$, where $f$ stands for the radiative
efficiency\footnote{Assuming that 
the electrons are fast cooling; to be verified in the next section.}. 
Combined with the energy density of the blob (\ref{eblob}), the typical dimension of the blob is
\be
\tilde{l}=(E_{\rm em}/\Gamma_{\rm em}\tilde{e}_{\rm em})^{1/3}\sim 10^{14}
\frac{L_{\rm f,47}^{1/3}t_{\rm f,300}^{1/3}r_2^{2/3}}{\Gamma_{\rm
  j,1}^{1/3}\sigma_2^{1/2}f_{1/2}^{1/3} L_{\rm j,47}^{1/3}}\quad
{\rm cm},\label{lcom}
\ee
of the order of the size of the black hole.
In the last expression we have assumed a quasi-spherical blob. The geometry of
the emitting material may be much more complex, depending on the details of 
the reconnection geometry and whether the plasma radiates before it leaves
the reconnection region or further downstream. In the latter case, a
quasi-spherical emitter is more likely.

The emitting region cannot be arbitrarily large.
Causality arguments for the emitting region limit its size  to be smaller than 
$l'<\Gamma_{\rm em}ct_{\rm f}=9\times 10^{14} \Gamma_{\rm
  j,1}\sigma_2^{1/2}t_{\rm f,300}$ cm
(e.g. Begelman et al. 2008). This can be cast as a ``causality'' constraint to the bulk
Lorentz factor of the jet (using also eq. (\ref{lcom})),
\be
\Gamma_{\rm j}>2 \frac{r_2^{1/2}L_{\rm f,47}^{1/4}}{t_{\rm f,300}^{1/2}
\sigma_2^{3/4}f_{1/2}^{1/4}L_{\rm j,47}^{1/4}}.
\label{caus}
\ee

\section{Radiation mechanisms}

The blob contains energetic electrons that emit though the 
synchrotron-self-Compton (SSC) and, possibly, external inverse Compton
mechanisms. Here we explore the photon energies at which the different components are
emitted, the radiative efficiency and under which conditions TeV
emission can escape the source.

\subsection{Synchrotron-self-Compton emission}

The synchrotron emission of electrons with random Lorentz factor $\gamma_e$ 
takes place at observed energy
\be
\nu_{\rm syn}\simeq\Gamma_{\rm em}\gamma_e^2\nu_{\rm c}\simeq 1.2
L_{\rm j,47}^{1/2}\sigma_2^{3/2}f_{1/2}^2r_2^{-1}\quad {\rm keV},
\ee
where $\nu_{\rm c}$ is the electron cyclotron frequency.
The peak of the self-Compton emission appears at 
\be
\nu_{\rm SSC}\simeq \gamma_{\rm e}^2 \nu_{\rm syn}\simeq 120
L_{\rm j,47}^{1/2}\sigma_2^{5/2}f_{1/2}^4r_2^{-1}\quad {\rm GeV},  
\ee
with the scattering taking place in the Thomson limit.

While the synchrotron emission peaks in the soft X-ray band, the
Comptonized component appears in the $\sim 100$ GeV range.
Any high-energy tail in the electron distribution powers the
$\sim$TeV flares. For fast electron cooling (see below), 
the relative strength of the synchrotron and inverse Compton components depends on the ratio
of the magnetic energy density to the radiation energy density in
the emitting region. For magnetization of the downstream plasma (blob)
of order unity and $f\sim 0.5$, the $y$-parameter in the emitting blob 
is $y\sim f\sigma_{\rm em}^{-1}\sim 1$. The cooling
timescale for the electrons in the blob rest frame  is $\tilde{t}_{\rm cool}\simeq
5\times 10^8/(1+y)\gamma_e\tilde{B}_{\rm em}^2$ s or
\be 
\tilde{t}_{\rm cool}\simeq 1.8\times 10^2 \Big(\frac{2}{1+y}
\Big)\frac{\Gamma_1^2r_2^{2}}
{L_{\rm j,47}\sigma_2^{1/2}f_{1/2}}\quad {\rm s}.
\label{tcool}
\ee

To evaluate whether there is efficient TeV emission, the cooling time is to be compared
to the time it takes for the blob to be slowed down after it is ejected in the
jet. The blob interacts with the rest of the jet plasma, and
rarefaction and shock waves form that propagate in the blob with a speed $\sim c/2$ 
(for the relativistically hot, $\sigma_{\rm em}\simless 1$ blob under consideration). 
The blob slows down on a timescale $\tilde{t}_{\rm s}\sim 2\tilde{l}/c$ as measured in its rest
frame. Efficient TeV radiation takes place when $\tilde{t}_{\rm
  cool}<\tilde{t}_{\rm s}$, which can be cast as (using eqs. \ref{lcom} and \ref{tcool})
\be
\Gamma_{\rm j}<48\Big(\frac{1+y}{2}\Big)^{3/7}
\frac{L_{\rm f,47}^{1/7}t_{\rm f,300}^{1/7}
f_{1/2}^{2/7}L_{\rm j,47}^{2/7}}{r_2^{4/7}}.
\label{effic}
\ee

The last expression implies that the blob can move with $\Gamma_{\rm em}\sim
\Gamma_{\rm j}\sigma^{1/2}\sim 100$ or larger
(for $\sigma\sim 100$) and still be in a fast SSC
cooling regime, in contrast 
to expectations from uniformly moving jets (Begelman et al. 2008). This is a result
of the compression that takes place in the reconnection layer, allowing 
for higher magnetic energy density and shorter cooling timescale compared to those
of a jet that moves uniformly with $\Gamma_{\rm j}\sim 100$.

Pair production on synchrotron photons might prevent $\sim 1$ TeV photons from escaping the blob. 
The energy density of synchrotron photons is at most that of the heated electrons
$\tilde{e}_{\rm syn}\sim f \tilde{e}_{\rm em}\sim 6 L_{\rm j,47}f_{1/2}r_2^{-2}
\Gamma_1^{-2}\quad {\rm erg/cm^3}$ while the bulk of the emission takes place 
at energy $\tilde{\nu}_{\rm syn}\sim \gamma_{\rm e}^2\nu_{\rm c} \sim 12
L_{\rm j,47}^{1/2}\sigma_2f_{1/2}^2r_2^{-1}\Gamma_{\rm j,1}^{-1}\quad {\rm eV}$
(in the rest frame of the blob). The comoving number density of photons at the
peak of the synchrotron emission is
\be
\tilde{N}_{\rm syn}^{\rm peak}\simeq \tilde{e}_{\rm syn}/h\tilde{\nu}_{\rm syn}=3.2\times 10^{11} 
\frac{L_{\rm j,47}^{1/2}}{\sigma_2f_{1/2}
r_2\Gamma_1} \quad {\rm ph/cm^3}.\label{Npeak}
\ee  

Most of the target photons that annihilate with the $\sim$1 TeV
$\gamma$-rays are close to the pair-creation threshold, i.e. at 
$\tilde{\nu}_{\rm target}\sim 0.6 \Gamma_{\rm em}\quad {\rm eV}=60 
\Gamma_1\sigma_2^{1/2}$ eV, typically above 
the bulk of the synchrotron emission. Assuming conservatively a fast cooling spectrum
$f_\nu\sim \nu^{-1}$ between $\tilde{\nu}_{\rm syn}$ and $\tilde{\nu}_{\rm target}$, 
the number density of the target photons
is a factor $\sim(\tilde{\nu}_{syn}/\tilde{\nu}_{\rm target})$ smaller than
that at the peak frequency.
Applying the last correction factor to eq.~(\ref{Npeak}), we estimate
\be
\tilde{N}_{\rm syn}^{\rm target}=6.3\times 10^{10} \frac{L_{\rm j,47}f_{1/2}}{\sigma_2^{1/2}
r_2^{2}\Gamma_1^{3}} \quad {\rm ph/cm^3}.  
\ee
As the TeV $\gamma$-rays cross the blob, they encounter optical 
depth to pair creation of
\be
\tau_{\gamma\gamma}\simeq \sigma_T\tilde{N}_{\rm syn}^{target}\tilde{l}/5=0.83  
\frac{L_{\rm f,47}^{1/3}t_{\rm f,300}^{1/3}L_{\rm j,47}^{2/3}f_{1/2}^{2/3}}{\sigma_2
r_2^{4/3}\Gamma_1^{10/3}}.
\ee
From the last expression, we obtain the condition that $\tau_{\gamma\gamma}<1$ is
satisfied for jet bulk Lorentz factor
\be
\Gamma_j>9 \frac{L_{\rm f,47}^{1/10}t_{\rm f,300}^{1/10}L_{\rm j,47}^{1/5}
f_{1/2}^{1/5}}{r_2^{2/5}\sigma_2^{3/10}}.
\label{transp}
\ee
This limit is much less stringent than the one ($\Gamma_{\rm j}>50$) found in
homogeneous jet models (Begelman et al. 2008; Mastichiadis \& Moraitis 2008).
This moderate value of $\Gamma_{\rm j}$ can be easily reconciled with
values inferred by the radio observations (e.g. Foschini et al. 2007) and 
unification schemes for AGN jets.   

We conclude that there is a reasonably wide range of Lorentz factors of the jet, 
bounded by the expressions (\ref{effic}) and (\ref{transp}), for which the 
SSC mechanism emits efficiently in the $\sim 100$ GeV-TeV range and the emission escapes
the source. The ``causality'' constraint for the Lorentz factor of the source
(\ref{caus}) is also satisfied.

\subsection{External inverse Compton emission}

External inverse Compton (EIC) may also contribute to the $\gamma$-ray emission and to opacity for
the $\gamma$-rays. It is, however, not necessary for efficient TeV emission in
our model. For the EIC mechanism to dominate the SSC, the
energy density of external soft photons must exceed that of the magnetic field
in the rest frame of the blob. The lab-frame energy
density of the external radiation must be
\be
U_{\rm soft}>6\times 10^{-4} \frac{L_{\rm j,47}f_{1/2}}{\Gamma_1^{4}\sigma_2
r_2^{2}} \quad {\rm erg/cm^3}.
\ee
This is too high to be attributed to the accretion disk (see also Begelman et
al. 2008). We cannot exclude, however, a powerful external source of soft
photons that is located in the vicinity of the hot blob. This source may
provide additional soft photons to be upscattered to the $\sim$TeV range.

\section{Statistics of flares}

In the ``jets-in-a-jet'' model discussed here, the emitting region moves
with large bulk $\Gamma_{\rm em}\sim 100$. The emission from the blob is beamed into
a narrow cone $\Delta\Omega_{\rm em}\sim 1/4\Gamma_{\rm em}^2\sim 2.5\times
10^{-5}\Gamma_{\rm j}^2\sigma_2$ 
of the sky and is directed at an angle $\theta\sim 1/\Gamma_{\rm j}$ 
with respect to the radial direction (see eq. \ref{angle}). The TeV emission
is beamed within the cone where the bulk of the jet emission takes place.

Assuming that  the jet opening angle is $\theta_{\rm j}\sim 1/\Gamma_{\rm j}$
and that the jet points at us,
the probability to see the emission from a single blob is $P\sim \Delta\Omega_{\rm
em}/\Delta\Omega_{\rm j} \sim 1/100 \sigma_2$.   
Observationally, the duty cycle of the flaring activity is low, maybe of the order of
$\sim 10^{-2}$. On the other hand, when it occurs, it is characterized by
several flares on $\sim 1$ hour timescales (e.g. Aharonian et al. 2007). 
So one needs to account for flares repeating on timescales of $t_{\rm rep}\sim 10^3 t_{\rm
  f, 300}$ s. 

If the emitting blobs are oriented randomly (in the rest frame of
the jet), then correcting for the blobs that are not emitting towards the
observer, the rate of dissipation events in the jet during the flaring
activity is $\sim  1/Pt_{\rm rep}\sim 0.1 \sigma_2t_{\rm f,300}^{-1}$ s$^{-1}$. 
The rate of dissipation of energy in the blobs corresponds to a significant fraction
of the jet power. Every blob contains lab-frame energy
$E_{\rm em}=1.5\times 10^{45}L_{\rm f,47}t_{\rm
  f,300}\Gamma_1^{-2}\sigma_2^{-1} f_{1/2}^{-1}$ erg (see Sect. 2.2.1).
The rate of dissipation in the jet is $L_{\rm diss}=E_{\rm em}/Pt_{\rm rep}\simeq 1.5\times
10^{44} L_{\rm f,47}\Gamma_1^{-2} f_{1/2}^{-1}$ erg/s. The
(corrected for beaming) luminosity of the jet is $L_{\rm j}^{cor}\simeq L_{\rm
j}\theta_{\rm j}^2/4\sim  L_{\rm j}/4\Gamma_{\rm j}^2 \simeq 2.5\times 10^{44}
 L_{\rm j,47}\Gamma_{1}^{-2}$ erg/s. The dissipated fraction of the jet
 luminosity is of order unity: $L_{\rm diss}/ L_{\rm j}^{cor}\simeq 0.6  L_{\rm f,47} L_{\rm
 j,47}^{-1}f_{1/2}^{-1}$. 
   
On the other hand, it is just as likely that the short-time 
variability is produced by intrinsic instabilities (e.g., tearing) 
of one large reconnection region. In this case, one large (1 hour-long)
flare involves the ejection of several individual plasmoids. 
Because their motion is controlled by the large-scale magnetic field, 
the directions of these blobs will no longer be random, but instead
will be strongly correlated with each other; this will significantly
lessen the flare energetics requirements.

Although we have focused on the fastest evolving flares in blazars,
which are the most constraining for the model, the same mechanism may be 
responsible for the observed variability on longer ($\sim$hours) 
timescales. The longer timescales may be 
due to larger emitting regions and/or longer cooling timescales of the 
electrons. Both of these conditions are likely to be met at larger distances from the 
black hole. Maybe the shortest variability comes from the reconnection
regions associated with polarity inversions  of the magnetic field that threads
the black hole (which might naturally have a scale set by the size 
of the black hole) while the longer term variability could be associated with 
current-driven instabilities that develop in the jet at larger distance.

\section{Discussion/Conclusions}
\label{sec-conclusions}

In this paper, we propose a jets-in-a-jet model as a plausible source
of the TeV flares in Mrk 501 and PKS 2155--304. We postulate the existence of
blobs that move relativistically within the jet. These can result
in fast-evolving flares and an environment transparent to 
$\gamma$-rays even for a jet with moderate $\Gamma_{\rm j}\sim 10$,
much easier reconciled with bulk Lorentz factors inferred from pc-scale 
observations (Piner \& Edwards 2004; Giroletti et al. 2004;
 Piner, Pant \& Edwards 2008) than models invoking higher-$\Gamma$ jets.

The jets-in-a-jet can be realized in a Poynting flux-dominated flow
where a fraction of the jet luminosity is dissipated in reconnection events.
Material leaves the reconnection site at relativistic speed, as measured in
the jet frame, while the liberated energy can power
bright flares through the synchrotron-self-Compton mechanism\footnote{
A similar mechanism has been proposed by Giannios (2006) for the X-ray 
flares seen in the afterglow of gamma-ray bursts.}. 
The synchrotron component appears in the soft X-rays and the inverse Compton 
in the $\sim 100$ GeV-TeV range. Simultaneous X-ray observations may have 
already revealed indications of such a TeV--X-ray correlation (Albert et al. 2007).
The model prediction of simultaneous X-ray flares 
assumes that most of the X-rays come from the flaring sites, but there 
could be other, slowly-varying regions that produce X-rays as well, 
perhaps further out so the photon densities are lower and pair production 
is not too large.  For example, the X-rays could also be produced in 
magnetically-dominated regions that result in little TeV emission. 

The mechanism presented here has similarities to a mechanism
proposed for the variability of the prompt emission of gamma-ray bursts
(GRBs), i.e. that it can be enhanced by relativistic motions within the GRB
jet. Such motions may result from magnetic dissipation (Blandford 2002;
Lyutikov 2006a; 2006b) or relativistic turbulence (Narayan \& Kumar 2008).

Ghisellini et al. (2008) proposed to explain the rapid TeV variability of Mrk
501 and PKS 2155--304 via the localized magneto-centrifugal acceleration of 
beams of electrons. In their model, the particles stream along the magnetic field lines at very
small pitch angles, resulting in negligible synchrotron emission 
and ``orphan'' TeV flares. The presence or absence of simultaneous X-ray flares
may be used to discriminate between the two models.
 
Although the jet is  Poynting flux-dominated, this is not necessarily the case for the material
that leaves the reconnection region which powers the flares. The magnetization
in this region is expected to be much lower than that of the bulk of the jet.
Inferring the magnetization of the jet by modeling the TeV flares (as done
e.g. by Ghisellini \& Tavecchio 2008) may be misleading in the context of the
model presented here.

The energy that is dissipated during the flaring activity
is a significant fraction of that of the jet, indicating 
that we are dealing with an efficient dissipation event. For the parameters adopted here
( $\Gamma_{\rm j}\sim 10$, magnetization in the jet $\sigma\sim 100$),
the dissipation could take place near or just outside the Alfv\'en radius of a
jet that is ejected with an initial Michel magnetization parameter $\mu\sim
\Gamma_{\rm j}\sigma\sim 1000$ (Michel 1969; Begelman \& Li 1994). In this 
region, the flow may be particularly prone to (kink-type)
current-driven instabilities (e.g. Eichler 1993; Begelman 1998; Giannios \& Spruit 2007)
that are currently being investigated by 3-dimensional MHD simulations
(e.g., Moll, Spruit \& Obergaulinger 2008; McKinney \& Blandford 2009) 
and may provide a plausible trigger for magnetic dissipation. Alternatively, efficient dissipation may 
result from reversal of the polarity of the magnetic field that threads
the black hole. If the reversal takes place on the light crossing timescale of
the black hole $\sim R_{\rm g}/c$, then parts of the jet with antiparallel magnetic 
fields can collide at $R\sim \Gamma_{\rm j}^2R_{\rm g}\sim
100R_g$, dissipating Poynting flux through reconnection at the location of the collision.     

Since the jet under consideration is ejected with $\mu\sim 1000$, it can be, in
principle, accelerated to terminal Lorentz factors as high as $\Gamma_{\rm j}\sim 1000$.
This does nor appear to happen in blazars. On the other hand, gamma-ray
burst jets do attain these high Lorentz factors. If the jets
in the two sources are launched with similar magnetization, the difference in
the acceleration efficiency may be understood by the difference in the
confining external medium (Tchekhovskoy, McKinney \& Narayan 2008;
Komissarov et al. 2008). Recent relativistic MHD simulations (Komissarov et
al. 2008; see also Tchekhovskoy et al. 2008 for force-free 
simulations) show that if the pressure of the external medium provides 
a collimating funnel, the acceleration is efficient, in contrast to the case of
a less collimating external pressure. It is possible that the collapsing star
provides such external pressure to the GRB jet.  The absence of similar
confinement in blazars might account for the difference in the acceleration
efficiency.  

\section*{Acknowledgments}
We thank the referee, Gabriele Ghisellini, for insightful comments that
greatly improved the manuscript. DG acknowledges support from the Lyman Spitzer 
Jr. Fellowship awarded by the Department of Astrophysical Sciences at 
Princeton University. DU is supported by National Science Foundation Grant 
No.\, PHY-0215581 (PFC: Center for Magnetic Self-Organization 
in Laboratory and Astrophysical Plasmas). MCB acknowledges support from NASA, 
via a {\it Fermi Gamma-ray Observatory} Guest Investigator grant.

\end{document}